\documentclass[twocolumn,superscriptaddress,prb,10pt]{revtex4-1}
\usepackage{verbatim}
\usepackage{braket}
\usepackage{amsmath,amssymb}
\usepackage{tikz}
\usepackage{graphicx}
\usepackage{color}
\usepackage[colorlinks,bookmarks=false,citecolor=blue,linkcolor=red,urlcolor=blue]{hyperref}
\usepackage{times}

\begin{document}

\title{Entanglement and quantum transport in integrable systems}

\author{Vincenzo Alba}
\affiliation{International School for Advanced Studies (SISSA),
Via Bonomea 265, 34136, Trieste, Italy, 
INFN, Sezione di Trieste}

\date{\today}

\begin{abstract} 
Understanding the entanglement structure of out-of-equilibrium many-body 
systems is a challenging yet revealing task. Here we investigate the entanglement dynamics 
after a quench from a piecewise homogeneous initial state in integrable systems. This is the 
prototypical setup for studying quantum transport, and it consists in the 
sudden junction of two macroscopically different and homogeneous states. By exploiting the 
recently developed integrable hydrodynamic approach and the quasiparticle 
picture for the entanglement dynamics, we conjecture a formula for the 
entanglement production rate after joining two semi-infinite reservoirs, 
as well as the steady-state entanglement entropy of a finite subregion. 
We show that both quantities are determined by the quasiparticles created 
in the Non Equilibrium steady State (NESS) appearing at large times at the 
interface between the two reservoirs. Specifically, the steady-state entropy coincides 
with the thermodynamic entropy of the NESS, whereas the entropy production rate  
reflects its spreading into the bulk of the two reservoirs. 
Our results are numerically corroborated using time-dependent Density Matrix Renormalization 
Group (tDMRG) simulations in the paradigmatic XXZ spin-$1/2$ chain.

\end{abstract}


\maketitle

\section{Introduction}

\begin{figure}[t]
\includegraphics*[width=0.85\linewidth]{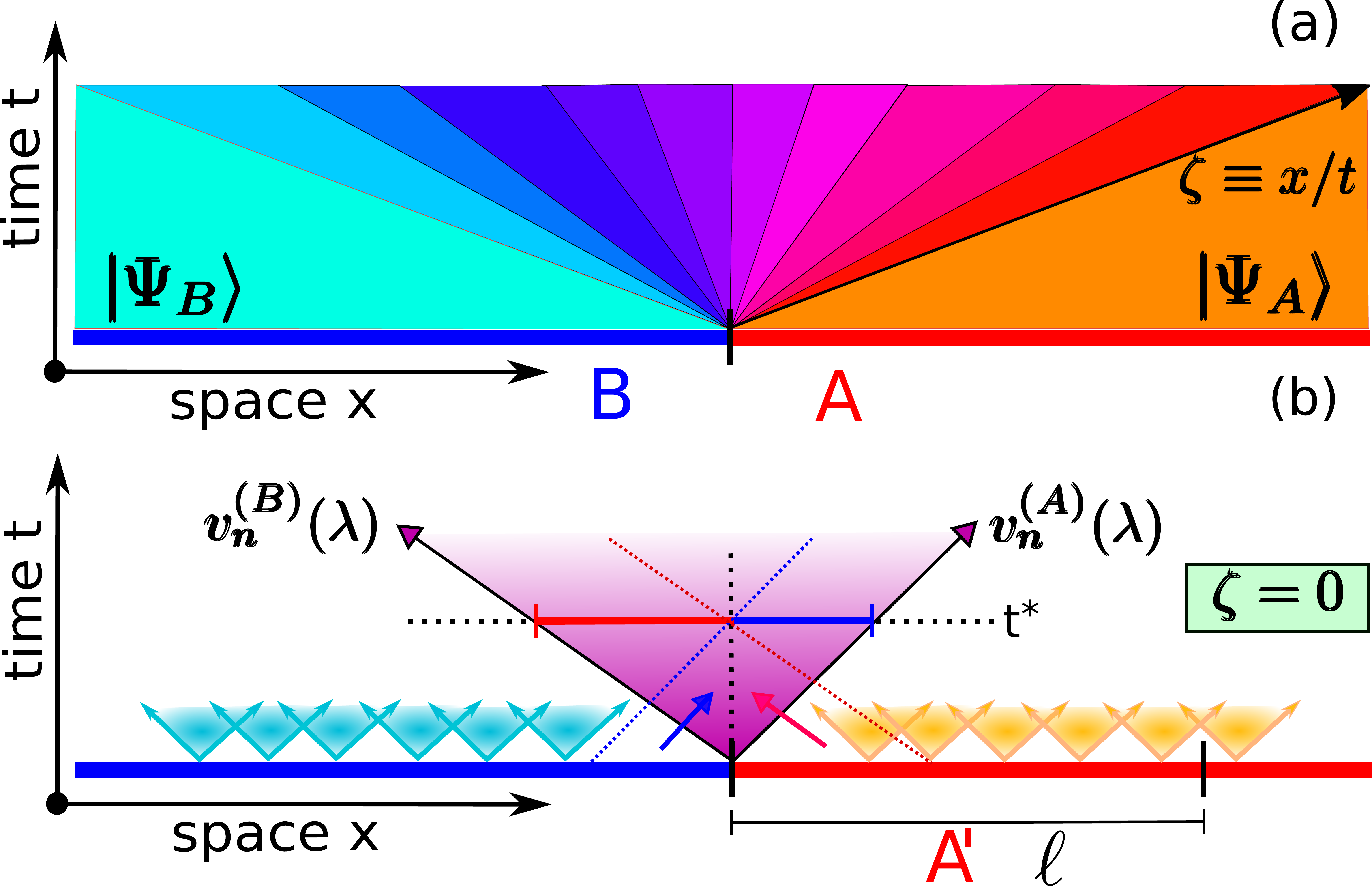}
\caption{ Entanglement dynamics after the quench from two homogeneous 
 chains  in integrable spin chains. (a) At $t=0$ two chains $A$ and 
 $B$ are prepared in the states $|\Psi_A\rangle$ and 
 $|\Psi_B\rangle$ and are joined. Dynamical 
 properties  at fixed $\zeta\equiv x/t$ are described by 
 a thermodynamic Bethe ansatz macrostate. (b) Quasiparticle picture for the 
 entanglement. Shaded cones denote quasiparticle 
 pairs. Different quasiparticles are produced in $A$ and $B$. 
 The entanglement production rate and its steady-state value are 
 determined by the macrostate with $\zeta=0$. The larger shaded region is 
 the associated lightcone (note the different velocities $v_n^{
 \scriptscriptstyle(A)}$ and $v_n^{\scriptscriptstyle(B)}$). 
 $n$ labels the quasiparticle families. We consider the 
 entanglement between $A$ and $B$, as well as that of a 
 finite region $A'$ of size $\ell$. 
}
\label{fig0}
\end{figure}

The quest for a complete understanding of entanglement spreading 
in out-of-equilibrium many-body system is a fruitful research theme. 
Recently, this became experimentally relevant, as 
it is now possible to measure the entanglement dynamics with 
cold atoms~\cite{islam-2015,kaufman-2016}. 
The best-known entanglement diagnostic tool is the von Neumann (entanglement) 
entropy $S\equiv-\textrm{Tr}\rho_A\ln\rho_A$, where $\rho_A$ is the 
reduced density matrix of a subsystem $A$ (see Figure~\ref{fig0} (a) for a 
one-dimensional setup).

A prominent out-of-equilibrium situation is that of the quench from two 
homogeneous initial states. This is a particular instance of 
{\it inhomogeneous} quantum quenches. The prototypical setup for spin chains is depicted in 
Figure~\ref{fig0} (a), and it consists in the sudden junction of two chains 
$A$ and $B$ that are prepared in two homogeneous {\it macroscopically} different  
quantum states $|\Psi_A\rangle$ and $|\Psi_B\rangle$. 
Subsequently, 
the state $|\Psi_A\rangle\otimes|\Psi_B\rangle$ is evolved in real time 
using a quantum many-body Hamiltonian $H$. Typically, a non-zero current 
arises between the two chains. The main focus so far has 
been on transport of local quantities, such as the local energy and 
local magnetization. Several techniques have been used, such as Conformal Field 
Theory~\cite{spyros-2008,bernard-2012,bhaseen-2015,allegra-2016,dubail-2017,dubail-2017a} (CFT), 
free-fermion methods~\cite{eisler-2009,de-luca-2013,sabetta-2013,eisler-2013,alba-2014,collura-2014,de-luca-2016,
eisler-2016,viti-2016,kormos-2017,perfetto-2017,vidmar-2017}, field theory methods~\cite{de-luca-2014,olalla-2014,biella-2016}, 
integrability~\cite{zotos-1997,zotos-1999,prosen-2011,prosen-2013,ilievski-2015}, and 
numerical techniques~\cite{fabian-2003,gobert-2005,langer-2009,karrasch-2012,karrasch-2013,vidmar-2017}. For integrable models 
a recent breakthrough~\cite{olalla-2016,bertini-2016} allows for an analytic treatment of transport 
problems using Thermodynamic Bethe Ansatz (TBA) techniques~\cite{doyon-2016,yoshimura-2016,de-luca-2016a,
doyon-2017,doyon-2017a,doyon-2017b,doyon-2017c,bulchandani-2017,bulchandani-2017a,ilievski-2017,doyon-2017d}. 

Surprisingly, as of now there are no exact results for the entanglement dynamics after inhomogeneous quenches. 
Notable exceptions are systems that can be 
mapped to CFTs in curved spacetime~\cite{allegra-2016,dubail-2017,dubail-2017a}. In 
this case it is established that $S$ grows logarithmically after the quench~\cite{dubail-2017}. 
%
%

In contrast, 
for {\it homogeneous} quenches a well-known quasiparticle picture~\cite{calabrese-2005} 
allows one to understand the entanglement dynamics  in terms of quasiparticles 
traveling ballistically, with opposite quasimomenta, through the system. The 
underlying idea is that only quasiparticles created at the same point in space 
are entangled. At a generic time $t$ the entanglement entropy $S(t)$ 
between a subregion $A$ and the rest is proportional to the number of 
quasiparticles emitted at the same point in space at $t=0$, and being  
at time $t$ one in subsystem $A$ and the other in $B$. More quantitatively, 
\begin{equation}
\label{quasi}
S(t)\propto 2t\!\!\!\!\int\limits_{\!2|v|t<\ell}\!\!\!\!d\lambda |v(\lambda)|f(\lambda)+
\ell\!\!\!\!\int\limits_{2|v|t>\ell}\!\!\!\!d\lambda f(\lambda), 
\end{equation}
where $f(\lambda)$ depends on the cross section for creating quasiparticles  with 
quasimomentum $\lambda$, and $v(\lambda)$ is their velocity. 
Eq.~\eqref{quasi} holds in the space-time scaling limit, i.e., $\ell,t\to\infty$ with 
fixed $t/\ell$. In many physical situations a maximum velocity $v_M$ exists, for instance, 
due to the Lieb-Robinson bound~\cite{lieb-1972}. Then, Eq.~\eqref{quasi} predicts 
a linear growth for $t\le \ell/(2v_M)$, followed by a volume-law $S\propto\ell$ at 
longer times. The validity of~\eqref{quasi} has been proven rigorously only for 
free-fermion models~\cite{fagotti-2008,ep-08,nr-14,coser-2014,cotler-2016,buyskikh-2016}, 
for which $f(\lambda)$ and $v(\lambda)$ can be determined analytically. The quasiparticle 
picture has been also confirmed in several numerical studies~\cite{de-chiara-2006,lauchli-2008,
kim-2013}, and using holographic methods~\cite{hrt-07,aal-10,aj-10,allais-2012,
callan-2012,ls-14,bala-2011,liu-2014}. Violations of the quasiparticle picture have 
been observed in CFT with large central charge~\cite{bala-2011,ab-13,asplund-2015,lm-15,kundu-2017,sagar-2017}. 
Remarkably, a framework to render~\eqref{quasi} predictive for generic integrable systems 
has been developed in Ref.~\onlinecite{alba-2016} (see Ref.~\onlinecite{mestyan-2017} for an application to 
the Hubbard chain). Specifically, for integrable systems $f(\lambda)$ coincides with 
the thermodynamic entropy associated with the post-quench steady state, while the 
quasiparticle velocities $v(\lambda)$ are those of the low-lying excitations 
around it.  

Here by combining a recent hydrodynamic approach for integrable 
systems~\cite{olalla-2016,bertini-2016} with the quasiparticle picture~\eqref{quasi}, 
we provide the first exact results for the entanglement dynamics after a global 
quench from a piecewise homogeneous initial state in generic integrable 
systems. Specifically, we focus on the steady-state 
entanglement entropy of a finite region $A'$ (see Figure~\ref{fig0} (b)), as well as 
the entanglement production rate between two semi-infinite reservoirs. Our main result is that both 
are determined by the physics of the Non-Equilibrium-Steady-State (NESS) 
(ray with $\zeta=0$ in Figure~\ref{fig0} (a)) appearing at the interface between 
the two reservoirs. The steady-state 
entanglement entropy coincides with the NESS thermodynamic entropy, whereas 
the entanglement production rate reflects the spreading of the NESS from the 
interface into the two reservoirs. 
 
The latter is relevant for quantifying the rate at which quantum information 
spreads. The study of the quantum information spreading is attracting enormous 
attention recently. For instane, in the holographic community, the 
focus has been on the so-called entanglement velocity $v_E$, 
which is defined as~\cite{hol-vel} $dS(t)/dt=v_E s_{th}\partial A$, with $s_{th}$ the thermodynamic 
entropy of the steady state, and $\partial A$ the length of the boundary between 
the subsystem of interest and the environment. For homogeneous quenches, $v_E$ 
has been investigated using holographic methods~\cite{hol-vel}, exact calculations 
in free models~\cite{cotler-2016,ff-vel}, and recently in the dynamics obtained 
from random unitary gates~\cite{ru-vel}. For inhomogeneous quenches, the 
information spreading started to be investigated only recently~\cite{erdmenger-2017}.

\begin{figure}[t]
\includegraphics*[width=0.93\linewidth]{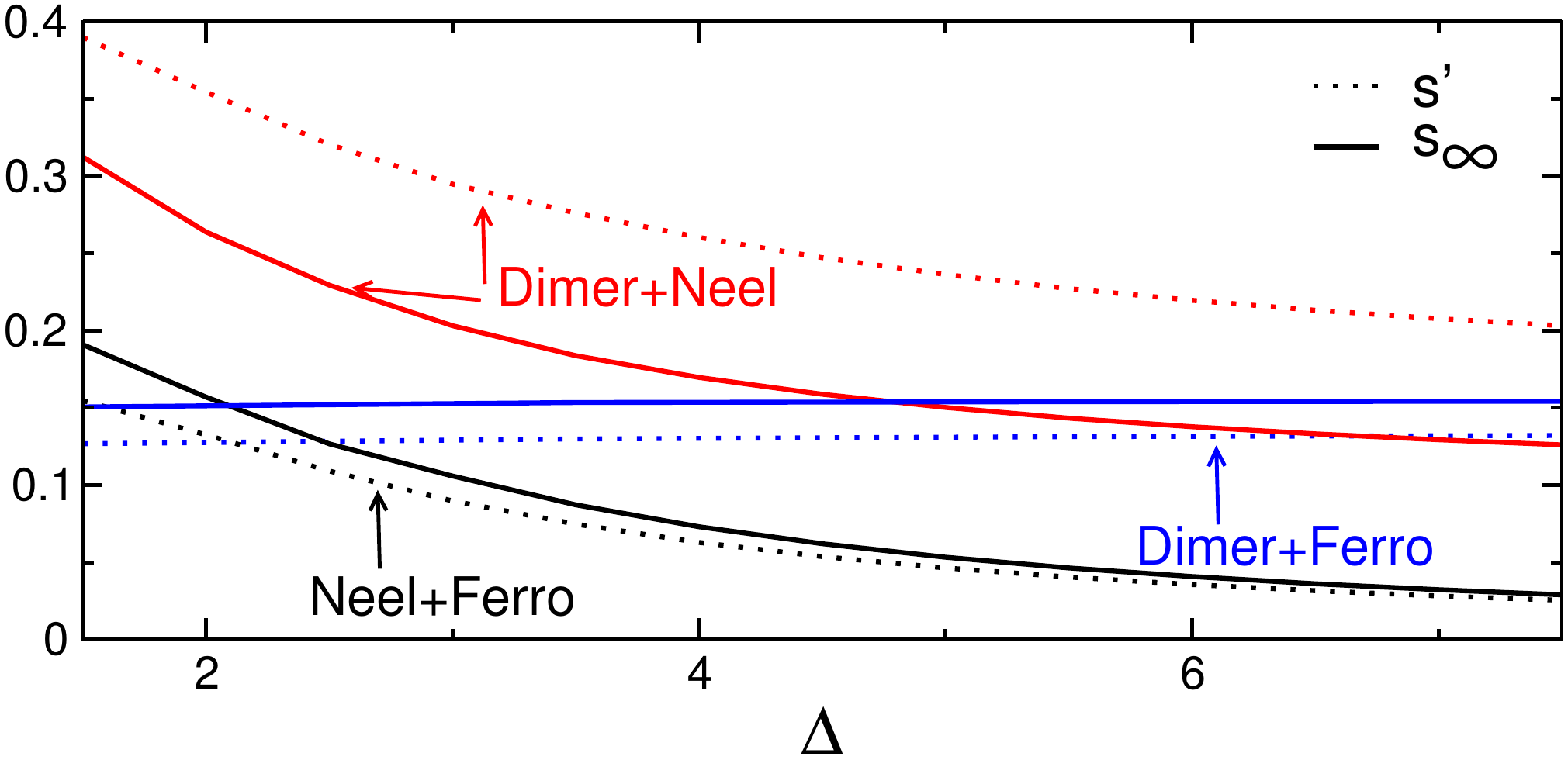}
\caption{ Entanglement dynamics after the quench from a piecewise homogeneous 
 initial state in the XXZ 
 chain: Theoretical predictions using the integrable hydrodynamics. 
 The steady-state entropy density $s_\infty\equiv\lim_{t\to\infty}S(t)/\ell$ 
 (full line) and the entanglement production rate $s'(t)\equiv dS(t)/dt$ 
 (dotted line) are plotted versus the chain anisotropy $\Delta$ for 
 several initial states. 
}
\label{fig1}
\end{figure}
%
\section{Entanglement via integrable hydrodynamics} 
The first key ingredient to derive our results is that the spectrum of integrable models 
exhibits families of stable quasiparticles. Typically, these are composite 
objects of elementary excitations. For spin chains, they correspond to bound states 
of magnons. The possible set of quasimomenta (rapidities) 
$\lambda$ that can be assigned to the quasiparticles are obtained by solving 
the so-called Bethe equations~\cite{taka-book} (see also Appendix~\ref{xxz-ba}). In the thermodynamic limit the 
rapidities form a continuum. Thermodynamic properties of integrable models 
are described by the particle densities $\rho_{n}(\lambda)$ and the hole densities 
$\rho^{\scriptscriptstyle (h)}_{n}(\lambda)$. The latter is the density of 
unoccupied rapidities. The total density 
is defined as $\rho^{\scriptscriptstyle(t)}_{n}\equiv\rho_{n}+\rho^{\scriptscriptstyle(h)}_{n}$. 
Here the subscript $n$ labels different quasiparticle families, and for spin chains is the 
size of the bound states. For free models in terms of quasimomenta 
$\rho_n^{\scriptscriptstyle (t)}=const$, reflecting that the quasimomenta are equally-spaced. 
Every set of $\rho_{n},\rho^{\scriptscriptstyle(h)}_{n}$ can be  
interpreted as a thermodynamic macrostate, which corresponds to an exponentially 
large number of microscopic eigenstates of the model. Their number is given 
as $e^{S_{YY}}$, with $S_{YY}$ the so-called Yang-Yang entropy~\cite{yang-1969} 
\begin{multline}
\label{syy}
S_{YY}=s_{YY}L=L\sum_{n=1}^\infty\int d\lambda[\rho_{n}^{\scriptscriptstyle(t)}\ln
\rho_{n}^{\scriptscriptstyle(t)}\\
-\rho_{n}\ln\rho_{n}-\rho_{n}^{\scriptscriptstyle(h)}\ln\rho_{n}^{\scriptscriptstyle(h)}]\\
\equiv L\sum_n\int d\lambda s_{YY}^{(n)}[\rho_{n},\rho_{n}^{\scriptscriptstyle(h)}]. 
\end{multline}
Similar to free models, $S_{YY}$ counts the number of ways of assigning the different 
rapidities $\lambda$ to the quasiparticles, compatibly with the densities $\rho_n,
\rho_{n}^{\scriptscriptstyle(h)}$. 

The second key ingredient in our approach is the integrable hydrodynamics 
framework~\cite{olalla-2016,bertini-2016}. 
Due to integrability, information spreads ballistically from the 
interface between $A$ and $B$. As a consequence, physical observables depend only 
on the combination $\zeta\equiv x/t$ (see Figure~\ref{fig0} (a)), with $x$ the distance 
from the interface between $A$ and $B$. For each fixed $\zeta$, dynamical 
properties of local and quasilocal observables are described by a thermodynamic 
macrostate, i.e., a set of densities $\rho_{\zeta,n},\rho_{\zeta,n}^{\scriptscriptstyle(h)}$. 
Each macrostate identifies a different Generalized Gibbs Ensemble~\cite{rigol-2008,polkovnikov-2011,calabrese-2005,rigol-2007,
cazalilla-2006,barthel-2008,cramer-2008,cramer-2010,calabrese-2011,cazalilla-2012a,calabrese-2012,sotiriadis-2012,
collura-2013,collura-2013a,fagotti-2013,fagotti-2014,kcc14,delfino-2014,sotiriadis-2014,ilievski-2015a,alba-2015,essler-2015,cardy-2015,
cardy-2015,sotiriadis-2016,bastianello-2016,vernier-2016,vidmar-2016,gogolin-2015,essler-2016,calabrese-2016,piroli-2017,piroli-2017a}. 
The macrostate with $\zeta=0$ is known as Non Equilibrium Steady State (NESS). 
The central result of Ref.~\onlinecite{olalla-2016,bertini-2016} is that 
$\rho_{\zeta,n}$ satisfy the continuity equation 
\begin{equation}
\label{cont}
[\zeta-v_{\zeta,n}(\lambda)]\partial_\zeta\vartheta_{\zeta,n}(\lambda)=0, 
\end{equation}
where $\vartheta_{\zeta,n}\equiv\rho_{\zeta,n}/\rho_{\zeta,n}^{\scriptscriptstyle(t)}$, 
together with the standard TBA equations~\cite{taka-book} 
\begin{equation}
\label{tba}
\rho^{(t)}_{\zeta,n}(\lambda)=a_n(\lambda)+\sum_{n=1}^\infty (a_{n,m}\star\rho_{\zeta,m})
(\lambda). 
\end{equation}
Here $a_n$ and $a_{n,m}$ are known functions of $\lambda$ (see Appendix~\ref{xxz-ba} for their 
expression for the XXZ chain), 
and the star symbols denotes the convolution $f\star g\equiv\int d\mu f(\lambda-\mu)g(\mu)$. 
Crucially, in~\eqref{cont} $v_{\zeta,n}$ are the velocities of the low-lying excitations around 
the macrostate $\rho_{\zeta,n}$, and fully encode the interactions (scatterings) between 
quasiparticles~\cite{bonnes-2014}. They can be calculated using standard TBA techniques~\cite{bonnes-2014} 
(see Appendix~\ref{velocities}). The solutions of~\eqref{cont} can be conveniently written 
as~\cite{bertini-2016} 
\begin{equation}
\label{sol}
\vartheta_{\zeta,n}(\lambda)=\theta_H(v_{\zeta,n}(\lambda)-\zeta)(\vartheta_n^B(\lambda)-\vartheta_n^A(\lambda))
+\vartheta^A_n(\lambda), 
\end{equation}
with $\theta_H(x)$ the Heaviside function, and $\vartheta_{\zeta,n}^{A(B)}$ the densities 
describing the steady states arising after the homogeneous quenches with initial states 
$|\Psi_A\rangle$ and $|\Psi_B\rangle$, respectively. Exact results for $\vartheta_n^{\scriptscriptstyle A(B)}$ are available for several 
quenches~\cite{caux-2013,wouters-2014A,pozsgay-2014A,bse-14,dwbc-14,bucciantini-15,
ac-16,pce-16,piroli-2016,mestyan-2015,brockmann-2014,iqdb-15,piroli-2016a,bpc-16,
caux-2016} (see also Appendix~\ref{steady}). 
%
\begin{figure}[t]
\includegraphics*[width=0.93\linewidth]{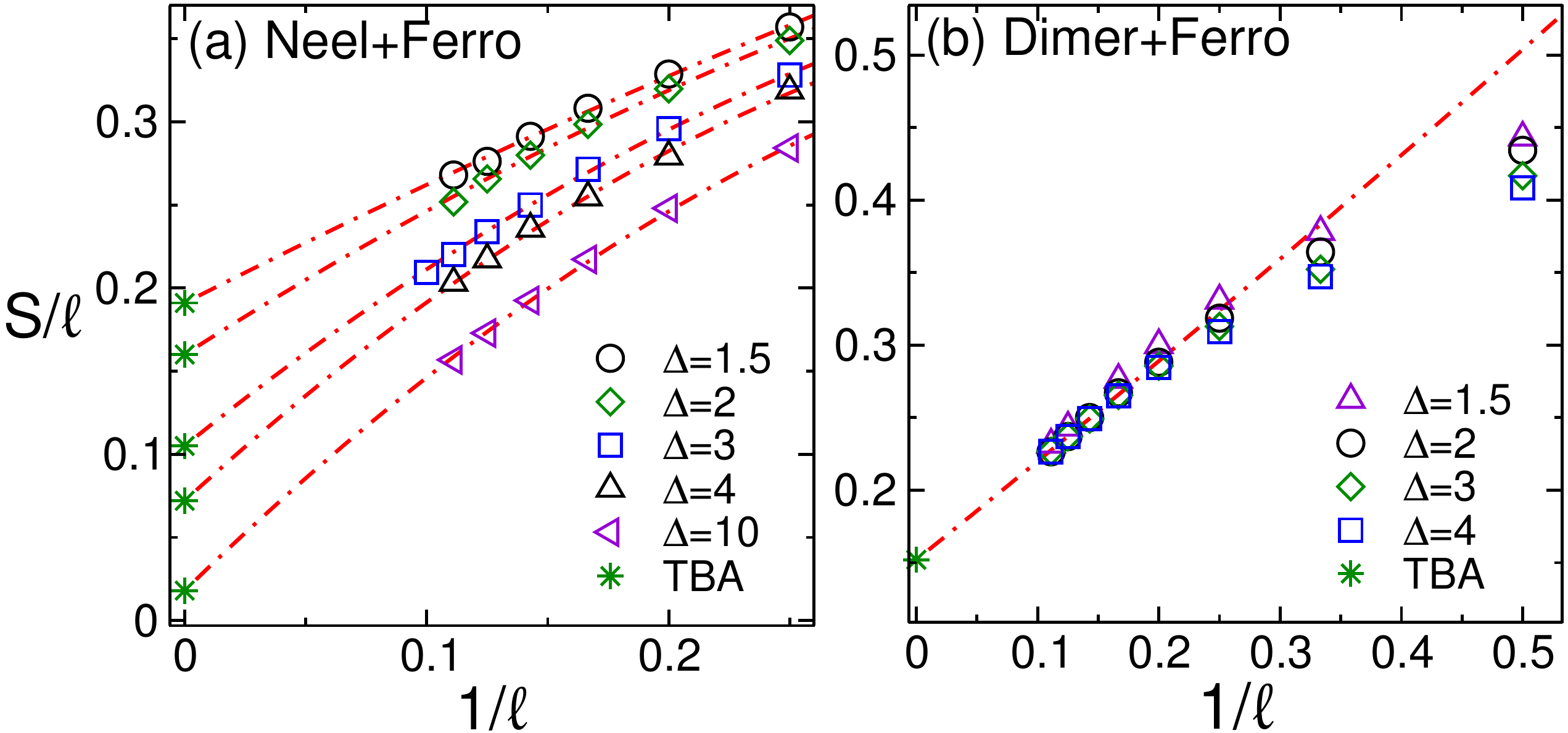}
\caption{ Steady-state entropy density after the quench 
 from a piecewise initial state in the XXZ chain. Results 
 are for the initial states $|N\rangle\otimes|
 F\rangle$ and $|MG\rangle\otimes|F\rangle$ (panel (a) and (b), 
 respectively). $S/\ell$ is plotted against $t/\ell$. The symbols are tDMRG 
 data at long times for different $\Delta$. The star symbols are the 
 Bethe ansatz results. Dash-dotted lines are fits to $S/\ell= s_\infty+
 a/\ell+b/\ell^2$, with $a,b$ fitting parameters, and $s_\infty$ the 
 Bethe ansatz results (Figure~\ref{fig1} (b)). 
}
\label{fig2}
\end{figure}
%
We now present our main results. We start discussing the steady-state 
entanglement entropy of a finite region $A'$ of length $\ell$ embedded in part $A$ and 
placed next to the interface with $B$ (see Figure~\ref{fig0} (b)). 
The steady-state entropy is obtained in the limit $\ell/t\to 0$, which identifies the macrostate 
with $\zeta=0$ (NESS). Since the spatial extension of the region described by this macrostate increases 
linearly with time, for $t\gg \ell$ the $\zeta=0$  macrostate is expected to 
describe the entire subsystem $A'$. Similar to Ref.~\onlinecite{alba-2016}, it is 
natural to conjecture that the entanglement entropy density $s_\infty\equiv\lim_{t\to\infty}S(t)/\ell$ 
becomes that of the Yang-Yang entropy of the NESS. The latter is the thermodynamic entropy of the 
GGE describing (quasi) local observables in space-time regions with $x/t\to0$. Using~\eqref{syy}, 
this implies  
\begin{equation}
\label{conj}
s_\infty=\sum_{n=1}^\infty\int d\lambda s^{\scriptscriptstyle(n)}_{YY}[
\rho_{\zeta=0,n},\rho_{\zeta=0,n}^{\scriptscriptstyle(h)}], 
\end{equation}
where $s^{\scriptscriptstyle(n)}_{YY}$ is calculated using the solutions of~\eqref{cont}\eqref{tba}. 

We now turn to the entanglement production rate. For homogeneous quenches, this corresponds to the 
slope of the linear term in ~\eqref{quasi}. Physically, Eq.~\eqref{quasi} means that $S(t)$ is determined by 
the total number of quasiparticles that at time $t$ crossed the interface between $A$ and $B$. 
It is natural to assume that the same applies  to the inhomogeneous case. 
After the quench, the different lightcones associated with different $\zeta$s  
start spreading from the interface between $A$ and $B$. For short times, all the 
types of quasiparticles remain confined within $A$, the only ones crossing the 
boundary between $A$ and the rest are the ones described by the $\zeta=0$ macrostate. 
At generic time $t$ their number is proportional to the width of the associated 
lightcone (see Figure~\ref{fig0} (b)). Equivalently, the entanglement 
growth at short times reflects the spreading of the NESS. 
Since the 
lightcone width increases linearly with time, one should expect the linear 
behavior $S(t)\propto s't$, with $s'\equiv dS(t)/dt$ the entanglement production rate given as  
\begin{equation}
\label{conj1}
s'=\sum_{n=1}^\infty\int d\lambda |v_{\zeta=0,n}|s_{YY}^{(n)}[\rho_{\zeta=0,n},
\rho_{\zeta=0,n}^{\scriptscriptstyle(h)}]. 
\end{equation}
Formally, Eq.~\eqref{conj1} is the same as that for the homogeneous quench conjectured in 
Ref.~\onlinecite{alba-2016}. However, here the quasiparticle 
velocities are not odd under parity, i.e., $v_n(\lambda)\ne-v_n(-\lambda)$, implying that 
the lightcone is not symmetric (in Figure~\ref{fig0}(b) $v_n^{\scriptscriptstyle(A)}$ and 
$v_n^{\scriptscriptstyle (B)}$ denote the different quasiparticle velocities in the two reservoirs). 
Similar to the homogeneous case the velocity of the entangling particles depends only on the local 
equilibrium state at large times~\cite{alba-2016} (here the NESS). Finally, we should mention 
that obtaining the full-time dynamics of the entanglement entropy is an arduous task, unlike for 
homogeneous quenches. The reason is that it requires reconstructing the 
quasiparticles trajectories. Precisely, since the quasiparticles velocity $v$ is finite, 
the typical time $t^*$ for the quasiparticles to travel through the subsystem is 
$t^*\sim\ell$, with $\ell$ the subsystem size. This implies that these 
quasiparticles should be described by $\zeta=\ell/t^*={\mathcal O}(1)$. A possible 
direction to determine the quasiparticle trajectories is to use the approach of 
Ref.~\onlinecite{doyon-corr}.

In the following we provide numerical evidence for~\eqref{conj} and~\eqref{conj1} considering 
the anisotropic spin-$1/2$ Heisenberg chain (XXZ chain). The Hamiltonian reads 
\begin{equation}
\label{xxz-ham}
H_{XXZ}=J\sum_{i=1}^L[S_i^xS_{i+1}^x+S_i^yS_{i+1}^y+\Delta S_i^zS_{i+1}^z]. 
\end{equation}
Here $S_i^{x,y,z}\equiv\sigma_i^{x,y,z}/2$ are spin-$1/2$ operators, 
$L$ the chain length and $\Delta$ the anisotropy parameter. We set $J=1$ 
in~\eqref{xxz-ham}, restricting ourselves to $\Delta>1$. 
The XXZ chain is the paradigm of Bethe ansatz solvable models~\cite{taka-book} (see 
Appendix~\ref{xxz-ba}). 

We consider the inhomogeneous quenches in which parts $A$ or $B$ are 
prepared in the N\'eel state $\left|N\right\rangle
\equiv(\left|\uparrow\downarrow\uparrow\cdots\right\rangle+\left|
\downarrow\uparrow\downarrow\cdots\right\rangle)/\sqrt{2}$, the Majumdar-Ghosh 
or dimer state $\left|MG\right\rangle\equiv[(\left|\uparrow\right\rangle+
\left|\downarrow\right\rangle)/\sqrt{2}]^{L/2}$, and the ferromagnetic 
state $|F\rangle\equiv\left|\uparrow\uparrow\cdots\right\rangle$. 
For all these cases the bulk densities $\vartheta_{n}^{\scriptscriptstyle 
A(B)}$ (cf.~\eqref{sol}) are known analytically (see Appendix~\ref{steady}). 
Also, for $\Delta>1$ the dynamics from states obtained by joining states with 
opposite magnetization~\cite{piroli-2017a} leads to diffusive or subdiffusive 
transport~\cite{ljubotina-2017,misguich-2017,stephan-2017}. In contrast, all the initial states 
considered here lead to ballistic transport (see Appendix~\ref{charge}). 

\begin{figure}[t]
\includegraphics*[width=1.\linewidth]{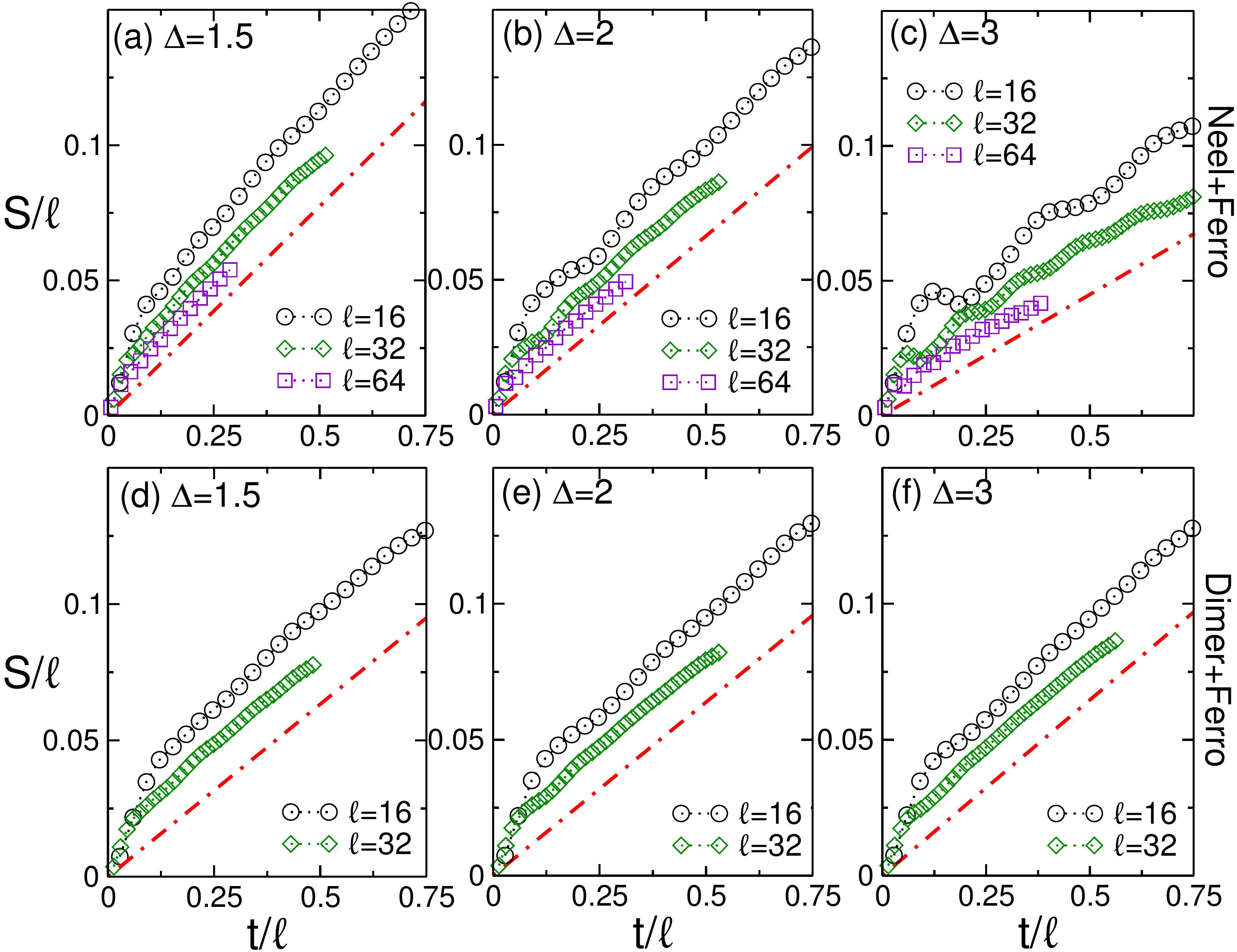}
\caption{ Entanglement dynamics after the quench from a piecewise homogeneous 
 initial state in 
 the XXZ chain: Entanglement production rate. Panels 
 (a,b,c) and (d,e,f) show tDMRG results for the initial states $|N\rangle
 \otimes|F\rangle$ and $|MG\rangle\otimes|F\rangle$, respectively. 
 Different symbols are for different $\Delta$ and $\ell\le 64$. 
 $S/\ell$ is plotted versus the rescaled time $t/\ell$. The dash-dotted 
 line is the theoretical result using Bethe ansatz (Figure~\ref{fig1}) 
 in the thermodynamic limit $\ell\to\infty$. 
}
\label{fig3}
\end{figure}

The strategy to use~\eqref{conj} and~\eqref{conj1} is to first solve the coupled 
systems of integral equations~\eqref{cont} and~\eqref{tba} for $\zeta=0$. The obtained densities 
$\rho_{\zeta=0,n},\rho_{\zeta=0,n}^{\scriptscriptstyle(h)}$ (some numerical results 
are shown in Appendix~\ref{neel-ferro}) are then substituted 
in~\eqref{syy}~\eqref{conj} and~\eqref{conj1}. Our results are summarized in Figure~\ref{fig1}.  
The Figure  shows the steady-state entropy $s_\infty$ 
(dotted lines) and the entanglement production rate  
$s'$ as a function of $\Delta$. For the quench with initial state $|N\rangle\otimes|
F\rangle$, both $s_\infty$ and $s'$ vanish for $\Delta\to\infty$, reflecting that the N\'eel state 
is the ground state of the XXZ chain in that limit. 
Interestingly, for both $|N\rangle\otimes|F\rangle$ and $|MG\rangle\otimes|F\rangle$, 
since the ferromagnet is an exact eigenstate of the XXZ chain at any $\Delta$, no quasiparticle production 
happens in subsystem $A$. As a consequence, $S(t)$ is fully determined by the quasiparticle 
transport from $B$ to $A$. However, $s_\infty$ and $s'(t)$ are smaller than the corresponding values for 
the homogeneous quench from the N\'eel state and the dimer state~\cite{alba-2016}. 
It is also interesting to observe that for these quenches only the quasiparticles with $n=1$ contribute 
in~\eqref{conj} and~\eqref{conj1}, i.e., the bound states contribution to the entanglement 
vanishes (see Appendix~\ref{neel-ferro}). This is not the case for the quench from 
$|MG\rangle\otimes|F\rangle$, for which all bound states contribute and quasiparticles are generated 
in both reservoirs. In this case both $s_\infty$ and $s'$ 
exhibit a rather weak dependence on $\Delta$ (see Figure~\ref{fig1}). 

\section{Numerical checks in the XXZ chain} 
We now turn to verify the theoretical predictions presented in Figure~\ref{fig1} using tDMRG 
simulations~\cite{white-2004,daley-2004,uli-2005,uli-2011,itensor}. 
We first focus on the steady-state entropy density $s_\infty$ of a finite block $A'$ 
of length $\ell$ (Figure~\ref{fig0} (b)). 
We consider only the quenches from the states $|N\rangle\otimes|F\rangle$ and $|MG\rangle\otimes|F\rangle$, 
as they are easier to simulate. In fact, in our tDMRG simulations we use the 
comparatively small value of the bond dimension $\chi\approx 75$. This allows us to obtain 
reliable results, provided that a space-time average of the data is performaed 
This surprising efficiency of tDMRG for transport-related quantities has been also observed in 
Ref.~\onlinecite{leviatan-2017}. 

Our tDMRG results in the regime $t\gg\ell$ are reported in Figure~\ref{fig2} (a) and (b), plotting $s_\infty$ versus 
$1/\ell$. The raw tDMRG data at any time after the quench are reported in Appendix~\ref{tdmrg}. To minimize 
the effects of oscillating (with the block size) scaling corrections we averaged the data 
for $t\gg\ell$. The results are for chains with 
$L=40$ sites, severals $\Delta$s (different symbols), and $\ell\le 10$. The star symbols 
are the Bethe ansatz results~\eqref{conj}. The dash-dotted lines are fits 
to $S/\ell=s_\infty+a/\ell+b/\ell^2$, with $s_\infty$ the thermodynamic limit results, 
and $a,b$ fitting parameters. Finite-size corrections are clearly visible 
for small $\ell$. However, for both initial states the numerical data are compatible in 
the thermodynamic limit with the Bethe ansatz results. Note that for the quench from 
$|MG\rangle\otimes|F\rangle$ the dependence on $\Delta$ for large $\ell$ is not visible within 
the numerical precision, as expected from Figure~\ref{fig1}. 

\begin{figure}[t]
\includegraphics*[width=1.\linewidth]{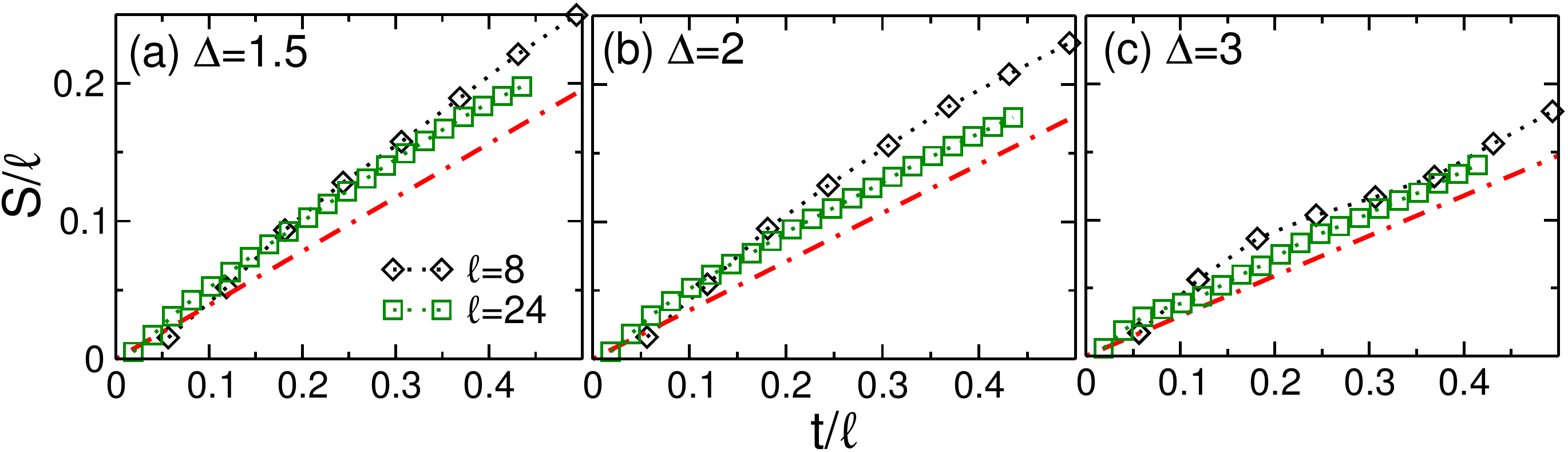}
\caption{ The same as in Figure~\ref{fig3} for the initial state 
 $|N\rangle\otimes|MG\rangle$. Notice the smaller subystems 
 sizes ($\ell\le 24$) as compared with Figure~\ref{fig3}. The 
 different panels are for different $\Delta$. The dash-dotted lines 
 are the Bethe ansatz results. 
}
\label{fig4}
\end{figure}

We now focus on the entanglement production rate in Figure~\ref{fig3}, again considering the 
quench from $|N\rangle\otimes|F\rangle$ (panels (a)-(c)) and $|MG\rangle\otimes|F\rangle$ (panels (d)-(e)), 
plotting $S/\ell$ versus $t/\ell$. The 
data are now for chains with $L\le 128$. We always consider the half-chain entropy, i.e., $A'=A$ 
and $A$ half of the chain (see Figure~\ref{fig0}). A crucial 
consequence is that only one boundary is present between $A'$ and the rest. The dash-dotted lines are 
the Bethe ansatz predictions (Figure~\ref{fig1} and~\eqref{conj1}). After an initial transient, 
in all cases $S(t)$ exhibits linear behavior. The slope of this linear increase is in agreement 
with the Bethe ansatz results, already for $\ell\approx 16$. A more systematic analysis of 
finite-size corrections is presented in Appendix~\ref{s-corr}. For the quench from $|MG\rangle\otimes|F\rangle$, 
again, one should observe the weak dependence on $\Delta$. Finally, in Figure~\ref{fig4} we 
show the results for the initial state $|N\rangle\otimes|MG\rangle$. Due to the large 
amount of entanglement, we can only provide reliable tDMRG data for $\ell\le 24$. tDMRG simulations 
are performed with $\chi=400$. To highlight finite-size 
and finite-time corrections we show data for $\ell=8$, which exhibit large deviations from 
the Bethe ansatz predictions. On the other hand, for $\ell=24$ the data are clearly  
compatible with~\eqref{conj1}. 

\section{Conclusions} We investigated the entanglement dynamics after the  
quenche from a piecewise initial state in integrable models. We conjectured an analytic formula for the 
entanglement production rate after joining two semi-infinite reservoirs as well as 
for the steady-state entropy of a finite subregion (\eqref{conj} and~\eqref{conj1}, 
respectively). Our work opens several promising 
new directions. 
For instance,  
it would be interesting to consider the inhomogeneous quench in which an 
integrability-breaking term acting at the interface between $A$ and $B$ is added to the Hamiltonian. 
Already in the case of a defect that preserves integrability this gives non trivial effects~\cite{bertini-2016}. 
It would be also 
enlightening to consider a time-dependent bipartition in which subsystem $A'$ moves away 
from the boundary. By appropriately tuning the speed at which $A'$ changes its position, 
it should be possible to change the thermodynamic macrostate  
governing the entanglement production. It should be possible to extend the 
method to treat the mutual information, as done for free fermions~\cite{eisler-2014,kormos-2017a}. 
Finally, it would be interesting to extend the quasiparticle picture to treat 
multipartite systems as in Ref.~\cite{savona}.

\begin{acknowledgments}
I am very grateful to Bruno Bertini for clarifying 
discussions on several aspects of the integrable hydrodynamic approach. I also acknowledge very fruitful discussions with 
Maurizio Fagotti and Pasquale Calabrese. This work was supported by the European  
Union's  Horizon  2020  research  and  innovation  programme under the Marie 
Sklodowoska-Curie grant agreement No 702612 OEMBS. 
\end{acknowledgments}

\appendix


\section{Bethe ansatz solution of the XXZ chain} 
\label{xxz-ba}

Due to the conservation of the total magnetization $S_z$, the eigenstates of the 
XXZ can be classified according to the total magnetization $S_T=\sum_i S_i^z$. 
Equivalently, one can use the total number $M$ of down spins as a good 
quantum number for the eigenstates. 

In the Bethe ansatz~\cite{taka-book} solution of the $XXZ$ chain, the eigenstates 
in the sector with $M$ down spins (particles) are identified by  
a set of $M$ rapidities $\lambda_j$, which are solutions of the so-called Bethe 
equations~\cite{taka-book} 
\begin{equation}
\label{be}
\left[\frac{\sin(\lambda_j+i\frac{\eta}{2})}{\sin(\lambda_j-i\frac{\eta}{2})}\right]^L=
-\prod\limits_{k=1}^M\frac{\sin(\lambda_j-\lambda_k+i\eta)}{\sin(\lambda_j-\lambda_k-i
\eta)},
\end{equation}
where $\eta\equiv\textrm{arccosh}(\Delta)$. 
Here we are interested in the thermodynamic limit $L,M\to\infty$ with $M/L$ fixed. 
Then, the solutions of the Bethe equations~\eqref{be} form string patterns in the 
complex plane. Rapidities forming a $n$-string can be written as 
\begin{equation}
\lambda^j_{n,\gamma}=\lambda_{n,\gamma}+i\frac{\eta}{2}(n+1-2j)+\delta^j_{n,\gamma}, 
\end{equation}
where $j=1,\dots,n$ denotes the different string 
components, $\lambda_{n,\gamma}$ is the ``string center'', and $\delta_{n,\gamma}^j$ 
are the string deviations. For the majority of the eigenstates 
of the XXZ chain, the string deviations are exponentially small, i.e., 
$\delta_{n,\gamma}^j={\mathcal O}(e^{-L})$ (string hypothesis). The 
$n$-strings describe bound states of $n$ down spins. The string centers 
$\lambda_{n,\gamma}$ are obtained by solving the Bethe-Gaudin-Takahashi (BGT) 
equations~\cite{taka-book} 
\begin{equation}
\label{bgt-eq}
L\theta_n(\lambda_{n,\alpha})=2\pi I_{n,\alpha}+\sum\limits_{(n,\alpha)
\ne(m,\beta)}\Theta_{n,m}(\lambda_{n,\alpha}-
\lambda_{m,\beta}). 
\end{equation}
For $\Delta>1$, one has $\lambda_{n,\gamma}\in[-\pi/2,\pi/2)$. Here we used that $\theta_n(
\lambda)\equiv2\arctan[\tan(\lambda)/\tanh(n\eta/2)]$. The scattering phases $\Theta_{n,m}
(\lambda)$ are given as 
\begin{multline}
\label{Theta}
\Theta_{n,m}(\lambda)\equiv(1-\delta_{n,m})\theta_{|n-m|}(\lambda)+2\theta_{
|n-m|+2}(\lambda)\\+\cdots+\theta_{n+m-2}(\lambda)+\theta_{n+m}(\lambda). 
\end{multline}
Each choice of the so-called BGT quantum numbers $I_{n,\alpha}\in\frac{1}{2}
\mathbb{Z}$ corresponds to a different set of solutions of~\eqref{bgt-eq}, i.e., 
to a  different eigenstate of the chain. The corresponding eigenstate energy $E$ 
and total momentum $P$ are obtained by summing over all the rapidities 
as $E=\sum_{n,\alpha}\epsilon_n(\lambda_{n,\alpha})$, and $P=\sum_{n,\alpha}z_n(
\lambda_{n,\alpha})$ with 
\begin{equation}
\label{eps}
\epsilon_n(\lambda)\equiv-\frac{\sinh(\eta)\sinh(n\eta)}{\cosh(n\eta)-\cos(2\lambda)}, 
\quad z_n(\lambda_{n,\alpha})=\frac{2\pi I_{n,\alpha}}{L}. 
\end{equation}
In the thermodynamic limit, one works with the rapidity densities. The root 
densities $\rho_{n,p}$  are formally defined as 
\begin{equation}
\rho_{n,p}(\lambda)\equiv\lim_{L\to\infty}  \frac1{L(\lambda_{n,\alpha+1}-\lambda_{n,\alpha})}. 
\end{equation}
It is also convenient to define the associated hole densities $\rho_{n,h}$, i.e., 
the density of unoccupied rapidities, and the total densities $\rho_{n,t}=
\rho_{n}+\rho_{n,h}$. The BGT equations~\eqref{bgt-eq} in the thermodynamic limit 
become a system of integral equations 
\begin{equation}
\label{tba1}
\rho^{(t)}_{n}(\lambda)=a_n(\lambda)+\sum_{n=1}^\infty (a_{n,m}\star\rho_{m})
(\lambda), 
\end{equation}
where we defined 
\begin{align}
\label{anm}
a_{nm}(\lambda)=&(1-\delta_{nm})a_{|n-m|}(\lambda)+2a_{|n-m|}(\lambda)\nonumber\\
 &+\ldots +2a_{n+m-2}(\lambda)+a_{n+m}(\lambda)\,,
\end{align}
with 
\begin{equation}
 a_n(\lambda)=\frac{1}{\pi} \frac{\sinh\left( n\eta\right)}{\cosh (n
 \eta) - \cos( 2 \lambda)}\,.
\end{equation}
%

\section{Macrostates for homogeneous quenches}
\label{steady}

Here we report the analytical results for the root densities $\vartheta^{
\scriptscriptstyle(A)}_n(\lambda)$ and $\vartheta^{\scriptscriptstyle(B)}_n(\lambda)$ (see 
the main text). We consider the homogeneous quenches from the N\'eel 
state and the Majumdar-Ghosh state. These densities characterize the thermodynamic 
macrostate in the bulk of the two systems $A$ and $B$, 
i.e., for $|\zeta|\to\infty$. We first define the densities $\eta_n\equiv
\rho_{n,h}/\rho_{n}$. In terms of $\eta_n$ one has $\theta_n=1/(1+\eta_n)$. 

For both the N\'eel state and the Majumdar-Ghosh state, the 
$\eta_n$ obey the recursive equation~\cite{ilievski-2015a,brockmann-2014}
\begin{align}
\label{eta-rec}
&\eta_n(\lambda)=\frac{\eta_{n-1}(\lambda+i\frac{\eta}{2})
\eta_{n-1}(\lambda-i\frac{\eta}{2})}{1+\eta_{n-2}(\lambda)}-1,\\
\label{rho-rec}
& \rho_{n,h}(\lambda)=\rho_{n,t}(\lambda+i\frac{\eta}{2})+\rho_{n,t}
(\lambda)-\rho_{n-1,h}(\lambda), 
\end{align}
where $\eta_0=0$ and $\rho_{0,h}=0$. 

For the N\'eel state one has~\cite{brockmann-2014} 
\begin{align}
\label{eta-neel}
& \eta_1=\frac{2[2\cosh(\eta)+2\cosh(3\eta)-3\cos(2\lambda)
\sin^2(\lambda)]}{[\cosh(\eta)-\cos(2\lambda)][\cosh(4\eta)-\cos(4\lambda)]}\\
\nonumber
& \rho_{1,h}=a_1\Big(1-\frac{\cosh^2(\eta)}
{a_1\pi^2\sin^2(2\lambda)+\cosh^2(\eta)}\Big), 
\end{align}
For the Majumdar-Ghosh one has~\cite{pozsgay-2014A,mestyan-2015}
\begin{align}
\eta_1=\frac{\cos(4\lambda)-\cosh(2\eta)}{\cos^2\lambda(\cos(2\lambda)
-\cosh(2\eta))}-1,
\end{align}
and 
\begin{equation}
\rho_{1,h}=a_1(\lambda)+\frac{1}{2\pi}(\omega(\lambda-i\eta/2)+\omega(\lambda+i\eta/2)), 
\end{equation}
with 
\begin{multline}
\omega(\lambda)=-\frac{\sinh(\eta)(-2+\cosh(\eta)+2\cosh(2\eta)}
{4(\cos(2\lambda)-\cosh(2\eta))^2}\\
+\frac{3\cosh(3\eta)+4\cos(2\lambda)(-\cosh(\eta)+\sinh^2(\eta))}
{4(\cos(2\lambda)-\cosh(2\eta))^2}. 
\end{multline}
%

\section{Velocities of entangling quasiparticles}
\label{velocities}

A crucial ingredient in the integrable hydrodynamic approach~\cite{olalla-2016,bertini-2016} 
and in the derivation of our results is the velocity of the low-lying excitations 
around the TBA macrostates $\rho_n,\rho_n^{\scriptscriptstyle (h)}$. Here we outline 
its derivation following the approach of Ref.~\onlinecite{bonnes-2014}. 
Given a generic thermodynamic macrostate identified by some densities $\rho_n,
\rho_n^{\scriptscriptstyle(h)}$, one can imagine 
of choosing among the eigenstates of the XXZ chain a representative of the macrostate.  
This would be identified by some BGT quantum numbers $I_{n,\alpha}^*$. Low-lying excitations around 
it can be constructed as particle-hole excitations. In each $n$-string sector, these correspond to 
the change $I^*_{n,h}\to I_{n,p}$, where $I_{n,p}(I^*_{n,h})$ 
is the BGT number of the added particle (hole). Since the model is interacting, 
this local change in quantum numbers affects {\it all} the rapidities obtained by solving the 
new set of Bethe equations. The excess energy 
of the particle-hole excitation can be written as~\cite{bonnes-2014} 
\begin{equation}
\label{ph-en}
\delta E_n=e_n(\lambda^*_{n,p})-e_n(\lambda^*_{n,h}). 
\end{equation}
Note that~\eqref{ph-en}  is the same as for free models apart from the dressing of the 
single particle energy. The change in the total momentum is obtained from~\eqref{eps} as 
\begin{equation}
	\delta P=z_n(\lambda^*_{n,p})-z_n(\lambda^*_{n,h}). 
\end{equation}
The group velocity associated with the particle-hole excitations is by definition  
\begin{equation}
\label{group-v}
v^*_n(\lambda)\equiv\frac{\partial e_n}{\partial z_n}=\frac{e'_n(\lambda)}{2\pi
\rho_{n}(1+\eta_n(\lambda))}, 
\end{equation}
where we used that~\cite{taka-book} $d z_n(\lambda)/d\lambda=2\pi\rho_{n,t}$, and 
we defined $e'_n(\lambda)=de(\lambda)/d\lambda$. Here $e'_n(\lambda)$ is 
obtained by solving an infinite system of Fredholm integral equations of the second 
kind~\cite{bonnes-2014}  
\begin{equation}
\label{tosolve}
e'_n(\lambda)+\frac{1}{2\pi}\sum\limits_{m=1}^\infty\int\!d\mu e'_m
(\mu)\frac{\Theta'_{m,n}(\mu-\lambda)}{1+\eta^*_m(\mu)}=
\epsilon'_n(\lambda),
\end{equation}
where $\Theta'_{n,m}(\lambda)\equiv d\Theta_{n,m}(\lambda)/d\lambda$ and 
$\epsilon'_n(\lambda)\equiv d\epsilon_n(\lambda)/d\lambda$  (cf. also~\eqref{Theta}
and~\eqref{eps}).


\begin{figure}[t]
\includegraphics*[width=1.\linewidth]{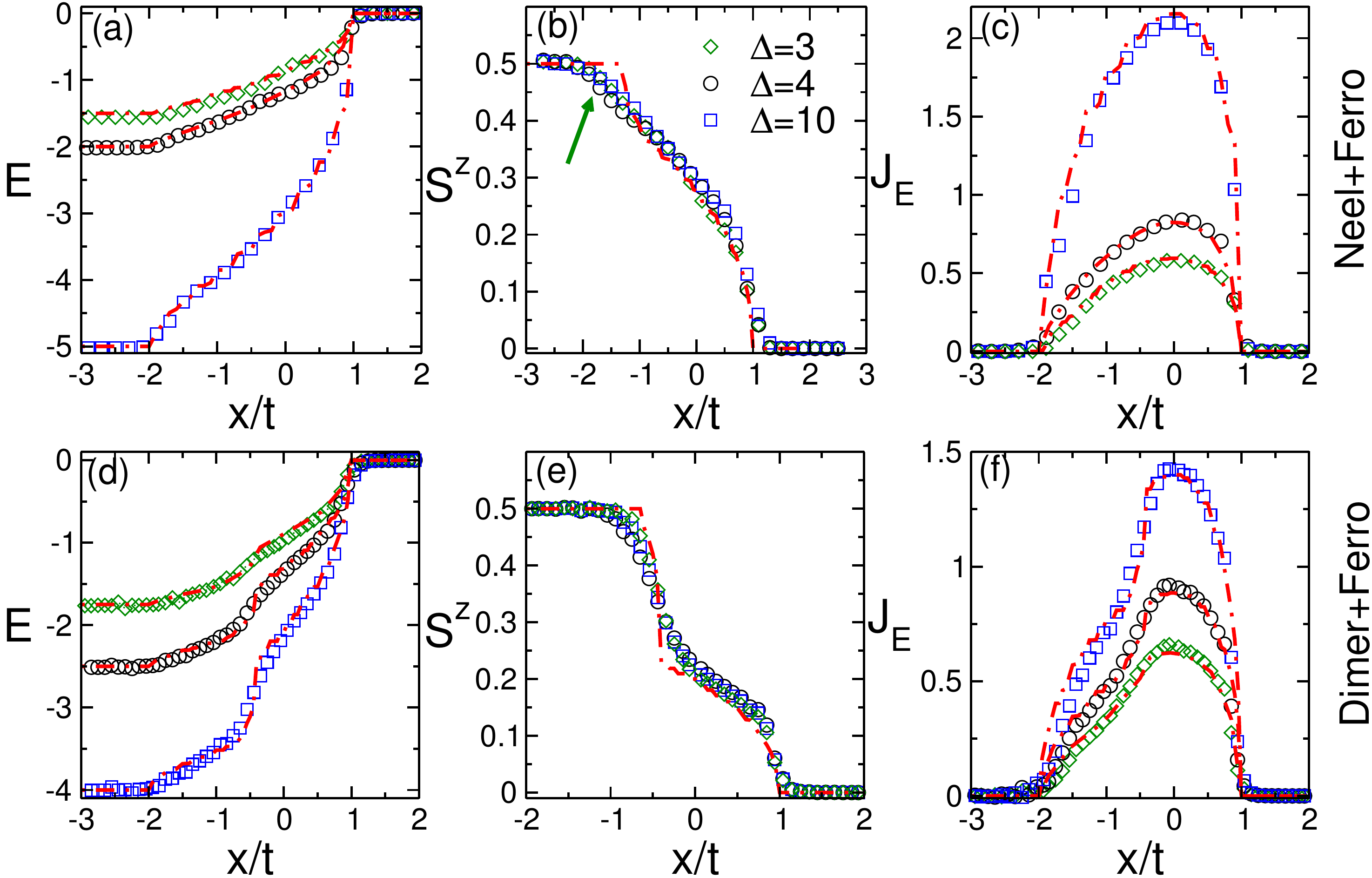}
\caption{ Quantum transport in the XXZ chain after the quench from 
 piecewise homogeneous initial state. 
 Panels (a,b,c) show results for the initial state $|N\rangle\otimes|F\rangle$. Panels (d,e,f) are for the 
 initial state $|MG\rangle\otimes|F\rangle$. The figure reports the local energy density $E$, the local magnetization 
 $S^z$, and the energy current $J_E$ plotted versus $\zeta=x/t$, with $x$ the position in the chain, measured
 from the interface between the two reservoirs. In all panels the 
 symbols are tDMRG data for a chain with $L=300$ and several values of $\Delta$. 
 To obtain smooth behavior a space-time average was performed in a window with 
 $\zeta=x/t\pm\epsilon$ and $\epsilon\approx 0.1$. 
 The dash dotted lines are the theoretical predictions using the Bethe ansatz. 
}
\label{fig6}
\end{figure}

\begin{figure}[t]
\includegraphics*[width=1.\linewidth]{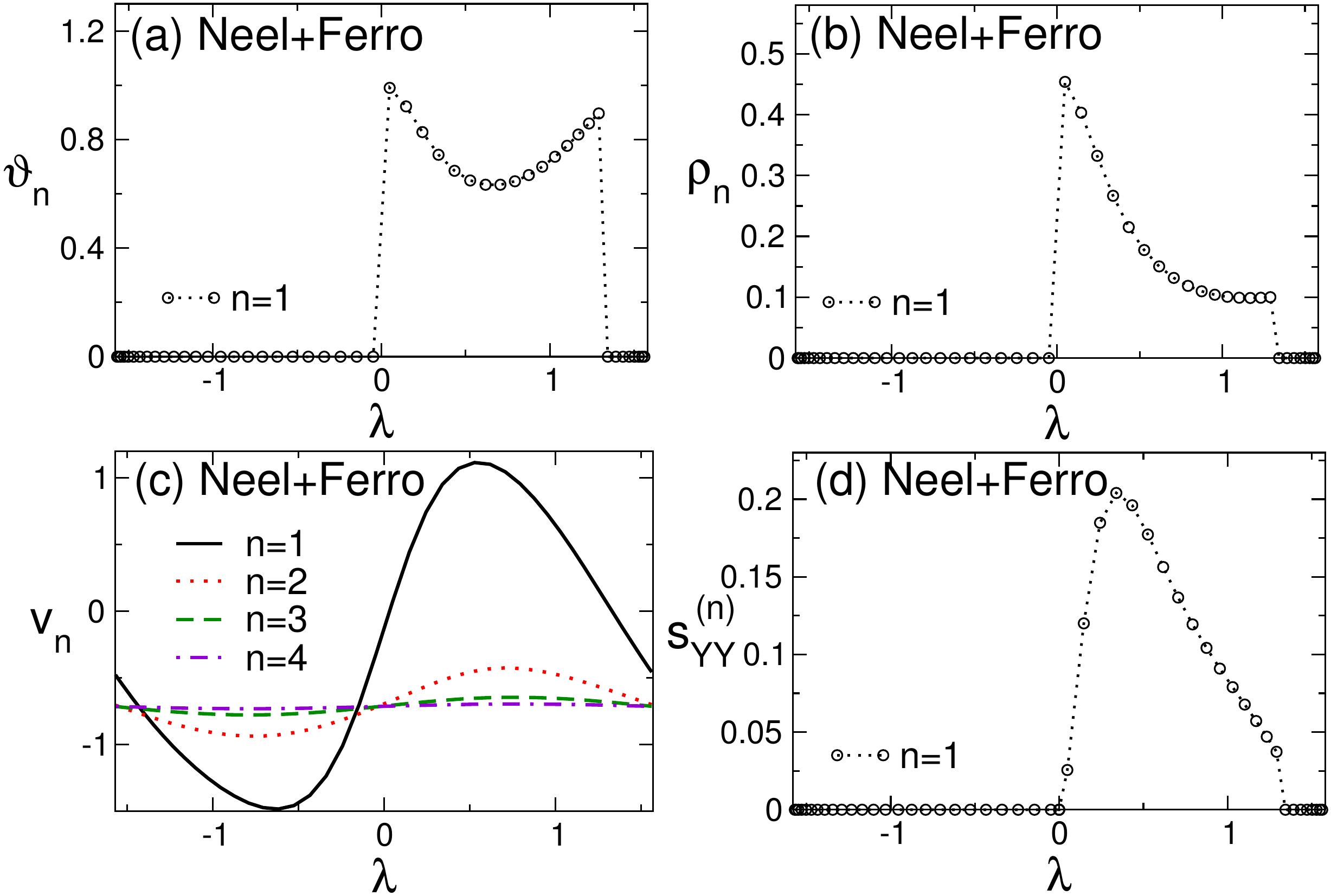}
\caption{ Integrable hydrodynamics approach for the quench from 
 the state $|N\rangle\otimes|F\rangle$. The results are for 
 the XXZ chain with $\Delta=2$. Panel (a) plots the densities 
 $\vartheta_n$ as a function 
 of the rapidity $\lambda$ for $\zeta=0$. 
 For $n>1$ one has $\vartheta_n=0$. Note that $\vartheta_1$ 
 is nonzero only for $\lambda>0$. Panel (b): Density 
 $\rho_n$ as a function of $\lambda$. Similar to (a) only 
 $\rho_1$ is nonzero for $\lambda>0$, whereas $\rho_n=0$ for 
 $n>1$. Panel (c): group velocities of the quasiparticles 
 excitations $v_n$ plotted as a function of $\lambda$. Different 
 lines correspond to different families of excitations (bound states). 
 Note that for $n>1$, one has that $v_n<0$ at any $\lambda$, whereas 
 $v_1$ changes sign at $\lambda=0$. Note also that $v_n$ have no 
 well defined parity, in contrast with homogeneous quenches 
 for which $v_n(\lambda)=-v_n(-\lambda)$. (d) Contributions to the 
 Yang-Yang entropy of the different quasiparticles families plotted 
 versus $\lambda$. Only $n=1$ gives nonzero contribution 
 for $\lambda>0$.  
}
\label{fig7}
\end{figure}

\section{Integrable hydrodynamics: Conserved charges and currents} 
\label{charge}

In this section we provide some numerical checks of the integrable hydrodynamics
approach~\cite{olalla-2016,bertini-2016}, focusing on the XXZ chain in the region with 
$\Delta>1$. We consider the quenches 
with initial states $|N\rangle\otimes|F\rangle$ and $|MG\rangle\otimes|F\rangle$. 
Using tDMRG simulations we investigate the dynamics of the local magnetization $S^z$, the 
local energy density $E$, and the energy current $J_E$. These are defined as~\cite{bertini-2016} 
\begin{align}
& S^z\equiv S_i^z\\
& E\equiv S_i^xS_{i+1}^x+S_i^yS_{i+1}^y+\Delta S_i^zS_{i+1}^z-\frac{\Delta}{4},\\
& J_E\equiv S_{i-1}^xS_i^zS_{i+1}^y-S_{i-1}^yS_i^zS_{i+1}^x\\\nonumber
&-\Delta S_{i-1}^zS_i^xS_{i+1}^y+\Delta S_{i-1}^zS_i^yS_{i+1}^x\\\nonumber
&-\Delta S_{i-1}^xS_i^yS_{i+1}^z+\Delta S_{i-1}^yS_i^xS_{i+1}^z. 
\end{align}
In the framework of the Bethe ansatz, these are written in terms of the root densities 
$\rho_{\zeta,n}(\lambda)$ as 
\begin{align}
\label{obs1}
& S_z=\sum_n n\int d\lambda\rho_{\zeta,n},\\
\label{obs2}
& E=\sum_n\int d\lambda\rho_{\zeta,n}\frac{\sinh\eta\sinh(n\eta)}{\cos(2\lambda)-\cosh(n \eta)},\\
\label{obs3}
& J_E=\sum_n\int d\lambda\rho_{\zeta,n}\frac{\sin(2\lambda)\sinh^2\eta\sinh(r\eta)}{
(\cos(2\lambda)-\cosh(n\eta))^2}. 
\end{align}
Note that all the observables are functions of $\zeta\equiv x/t$. The dependence on $\zeta$ is encoded 
in the densities $\rho_{\zeta,n}$~\cite{bertini-2016}. 

\begin{figure*}[t]
\includegraphics[width=1.\linewidth]{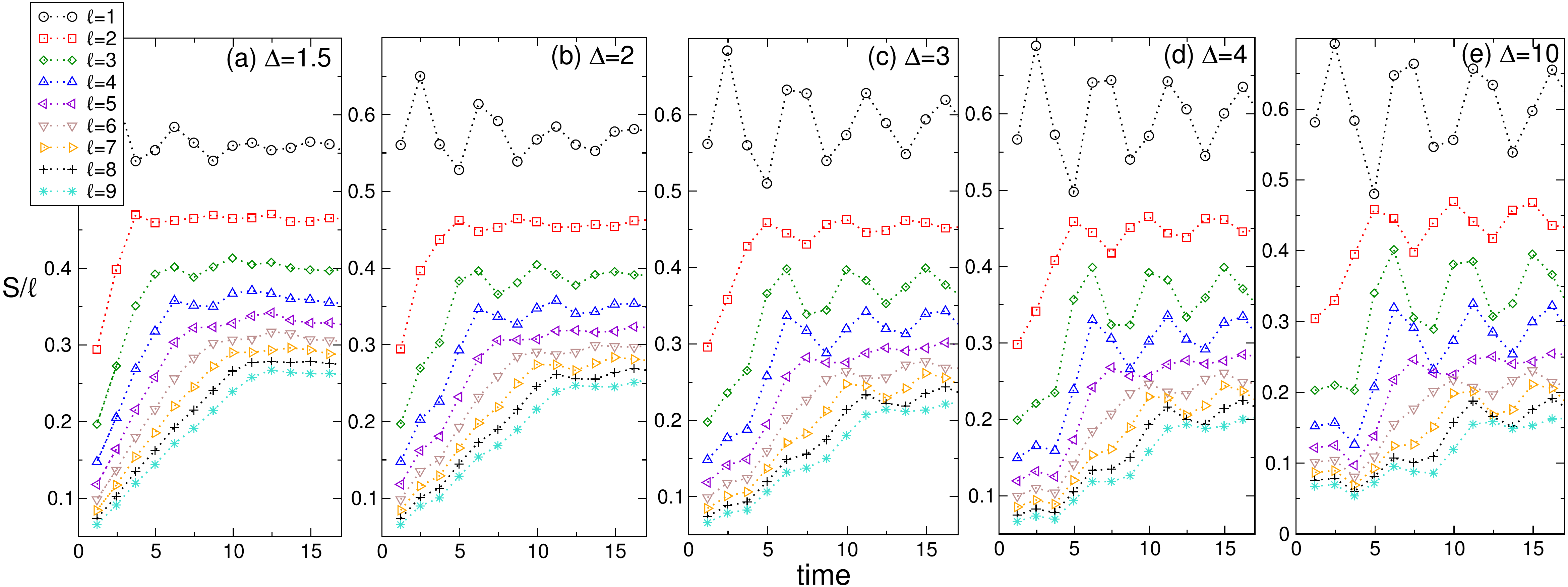}
\caption{ Entanglement dynamics after the quench from a piecewise homogeneous 
 initial state in the XXZ chain: Results 
 for initial $|N\rangle\otimes|F\rangle$. 
 The figure shows $S/\ell$ plotted versus time. The different panels are for 
 different values of $\Delta$. The symbols are tDMRG data for several values of the 
 subsystem length $\ell$ and for a chain with $L=40$ sites. 
}
\label{fig5}
\end{figure*}

The Bethe ansatz results~\eqref{obs1}\eqref{obs2}\eqref{obs3} are compared with tDMRG results in 
Figure~\ref{fig6}. The upper and lower panels show tDMRG data for the quench with initial 
states $|N\rangle\otimes|F\rangle$ and $|MG\rangle\otimes|F\rangle$, respectively. The data 
are for the XXZ chain with $L=300$ sites, up to times $t\lesssim 100$. The 
results in panels (a-c) are obtained using bond dimension $\chi\sim 20$. For the quench 
from the state $|MG\rangle\otimes|F\rangle$ (panels (e-f)) we 
used $\chi\sim 75$. In all panels the different symbols correspond to different values of $\Delta$. 
In order to remove spatial and temporal oscillations, we performed 
a spatio-temporal average. Specifically, for each fixed $\zeta$ the results in the Figure are obtained by averaging 
the data in a window $\zeta\pm\epsilon$ with $\epsilon\approx 0.1$. The dash-dotted lines are the theoretical 
results~\eqref{obs1}\eqref{obs2}\eqref{obs3}. For $|\zeta|\gg1$ all the observables become $\zeta$ independent. 
This happens in spatio-temporal regions with $|x|\gg v_Mt$, with $v_M$ the maximum velocity in 
the system. Remarkably, for both quenches, and for all values of $\Delta$, the tDMRG data are in 
very good agreement with the Bethe ansatz. Note, however, that the theoretical predictions 
exhibits some cusp-like behaviors (see for instance the arrow in 
panel (b)). These are expected and have been discussed in Ref.~\onlinecite{bb}, 
although they are not well reproduced by the tDMRG results. It is natural to expect that 
these deviations should be attributed to the finite $\chi$ and to the finite-time and finite-size 
effects.

\section{Neel-Ferro quench: Bethe ansatz results} 
\label{neel-ferro}

In this section we provide some details on the Bethe ansatz results 
for the quench from the state $|N\rangle\otimes|F\rangle$. 
Our results are summarized in Figure~\ref{fig7}. 

Specifically, panels (a) and (b) show the densities $\vartheta_{\zeta,n}$ 
and $\rho_{\zeta,n}$, respectively. The results are for $\zeta=0$, 
which identifies the relevant macrostate to describe the entanglement dynamics after the 
quench (the subscript $\zeta$ in $\vartheta_{\zeta,n}$ and $\rho_{\zeta,n}$ 
is omitted in the Figure). First, it is interesting to observe 
that only $\vartheta_{\zeta,1}$ and $\rho_{\zeta,1}$ are nonzero. Moreover, 
$\vartheta_1$ is nonzero only for $\lambda>0$. Similar behavior is observed 
for $\rho_1$ (see panel (b)). 

The group velocities of the low-lying excitations around the macrostate with $\zeta=0$ 
are reported in Figure~\ref{fig7} (c). These are obtained by numerically 
solving~\eqref{tosolve} and using~\eqref{group-v}. As anticipated in 
the main text, one has $v_n(\lambda)\ne-v_n(-\lambda)$. This is in stark contrast with 
homogeneous quenches, where one has~\cite{alba-2016} $v_n(\lambda)=-v_n(-\lambda)$. 
As a consequence the lightcone of the entangling quasiparticles is not 
symmetric. It is also interesting to observe that for any 
$\lambda$, $v_n<0$ for $n>1$, which implies that there is no transport of bound 
states from $B$ to $A$ after the quench.  

Finally, in Figure~\ref{fig7} (e) we show the contributions to the Yang-Yang 
entropy density $s_{YY}^{\scriptscriptstyle(n)}$ (see its definition in the main 
text) of the quasiparticles. Clearly, 
only quasiparticles with $n=1$ and $\lambda>0$ contribute to the entropy. Together with 
the results in panel (c), this implies that the entanglement between the two subsystems 
is generated by the transport of particles from $B$ to $A$. 

\section{Steady-state entropy: DMRG data} 
\label{tdmrg}

In this section we discuss in more detail the numerical data for the steady-state 
entanglement entropy presented in Figure $3$ in the manuscript, i.e., for the quench from the state 
$|N\rangle\otimes|F\rangle$. We consider the entanglement entropy of a subsystem $A'$ of 
length $\ell$ placed next to the boundary between $A$ and $B$. To avoid boundary effects $A'$ 
is embedded in $A$, which is chosen larger than $A'$. 

Our tDMRG results are presented in Figure~\ref{fig5}. The data are obtained using standard 
tDMRG simulations with bond dimension $\chi\sim 300$ for the XXZ chain with $L=40$ sites 
and $t\sim 16$. To calculate the von Neumann entropy we employed the techniques 
described in Ref.~\onlinecite{ruggiero-2016}. 
The different panels correspond to different values of $1.5\le\Delta\le 10$. The figure 
shows $S/\ell$ plotted versus the time after the quench. In each panel the different 
symbols are for subsystems of different length $1\le \ell\le 9$. For each $\ell$ a linear 
increase, followed by a saturation, is observed. The saturation value decreases upon increasing  
$\Delta$, in agreement with the theoretical predictions. 
For $t\to\infty$, in the limit $\ell\to\infty$, the entropy density $S/\ell$ should 
converge to the Bethe ansatz predictions. However, strong finite-size 
effects are visible in the Figure, due to the small $\ell$ considered. Moreover, it is interesting 
to observe that upon increasing $\Delta$, the data exhibit strong oscillations with time. 
The data presented in Figure $3$ correspond to the time average of the tDMRG data 
in the time window $13\le t\le 16$. Finally, we should mention that similar results are obtained 
for the quench from the state $|MG\rangle\otimes|F\rangle$.

\section{Scaling corrections} 
\label{s-corr}

In this section we discuss in more detail the finite-size corrections of the tDMRG data in 
Figure $4$ in the manuscript. We focus on the quench from the state $|N\rangle\otimes|F\rangle$, 
for which the largest system subsystem sizes up to $\ell=64$ are available. We consider the 
scaling corrections to the entanglement production rate. Figure~\ref{fig7} 
shows the deviations $\delta S/\ell$ plotted versus the rescaled time $t/\ell$. The theory 
predictions are obtained using formula~\eqref{conj1}. The different curves in 
Figure~\ref{fig7} correspond to different sizes of subsystem $A$ (see Figure $1$ in the 
manuscript). Clearly, corrections are vanishing in the scaling limit $t,\ell\to\infty$ with 
fixed ratio $t/\ell$. Interestingly, the dependence of the corrections on $t/\ell$ becomes 
weaker upon increasing $\ell$. By comparing the data for $\ell=32$ and $\ell=64$, it is clear 
that corrections decay as $1/\ell$ in the thermodynamic limit. This is supported in the inset 
of Figure~\ref{fig7} showing the data for fixed $t/\ell\approx 0.1$ and $t/\ell\approx 0.27$ 
plotted versus $1/\ell$. We should mention that a similar behavior as $1/\ell$ has been observed 
for homogeneous quenches~\cite{alba-2016}. 

\begin{figure}[t]
\includegraphics*[width=1.\linewidth]{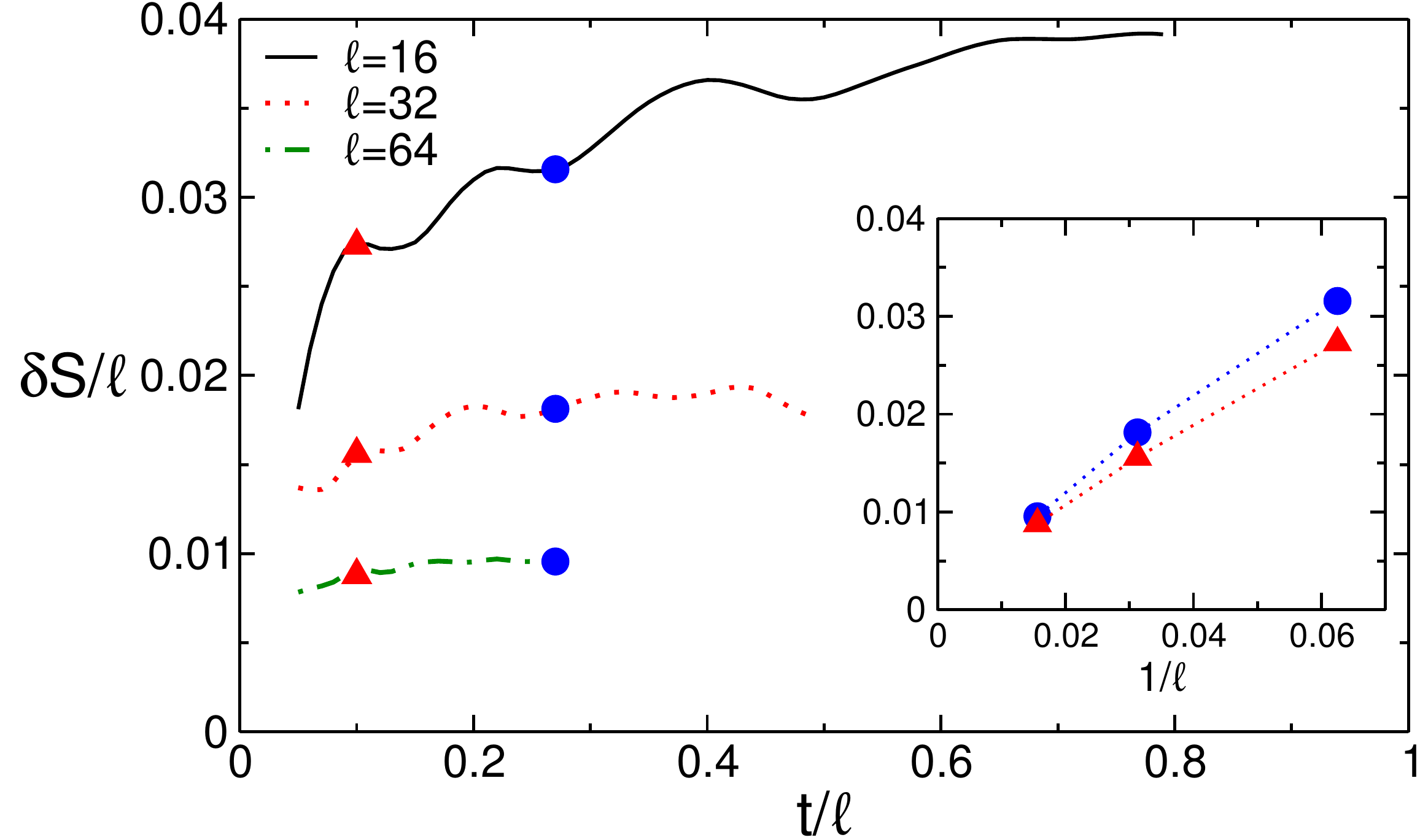}
\caption{ Entanglement production rate for the quench from the $|N\rangle\otimes|F\rangle$ quench: 
 finite-size corrections. In the main panel the curves are the deviations $\delta S/\ell$ from 
 the Bethe ansatz predictions for the quench with $\Delta=1.75$ (panel (a) in Figure $4$ in the 
 manuscript). Here $\delta S/\ell$ is plotted against the rescaled time $t/\ell$. The 
 different curves correspond to the different subsystem sizes $\ell=16,32,64$. Corrections are 
 clearly vanishing in the scaling limit. Inset: $1/\ell$ behavior of the scaling corrections. 
 Here $\delta S/\ell$ at fixed $t/\ell\approx 0.1$ and $t/\ell\approx0.27$ (data are marked with the 
 same symbols in the main panel) is plotted versus $1/\ell$.  
}
\label{fig7}
\end{figure}

\end{document}